\documentclass[letterpaper, 10pt, conference]{ieeeconf}  

\IEEEoverridecommandlockouts                              
\usepackage{amsmath,paralist,comment,color,amssymb}
\usepackage{cite}
\usepackage{supertabular}

 \DeclareMathOperator{\tr}{tr}

 \DeclareMathOperator{\swt}{swt}
\DeclareMathOperator{\wt}{wt}
\DeclareMathOperator{\Span}{span}

\newcommand{\C}{\mathbb{C}}
\newcommand{\F}{\mathbb{F}}

\newcommand{\sdual}{{\perp_s}}
\newcommand{\adual}{{\perp_a}}
\newcommand{\hdual}{{\perp_h}}

\newcommand{\ket}[1]{|#1\rangle}
\newcommand{\ds}{\displaystyle}
\newcommand{\scal}[2]{\langle #1\mid #2\rangle_s}

\newtheorem{theorem}{Theorem}
\newtheorem{corollary}[theorem]{Corollary}
\newtheorem{lemma}[theorem]{Lemma}
\newtheorem{remark}{Remark}

\newtheorem{example}{Example}

\title{\LARGE\bf { Subsystem Codes}
}

\author{Salah A. Aly, Andreas Klappenecker and Pradeep Kiran Sarvepalli
\thanks{This research was supported by NSF grant CCF-0622201, NSF CAREER award
CCF-0347310, and a TITF project.
}
\thanks{The authors are with the Department of Computer Science,
        Texas A\&M University, College Station, Texas 77843, USA 
        {\tt\small email:\{salah,klappi,pradeep\}@cs.tamu.edu}}%
}

\begin{document}

\maketitle
\thispagestyle{empty}
\pagestyle{empty}

\begin{abstract}
We investigate various aspects of operator quantum error-correcting
codes or, as we prefer to call them, subsystem codes.  We give various
methods to derive subsystem codes from classical codes.  We give a
proof for the existence of subsystem codes using a counting argument
similar to the quantum Gilbert-Varshamov bound.  We derive linear
programming bounds and other upper bounds. We answer the question 
whether or not there exist $[[n,n-2d+2,r>0,d]]_q$ subsystem
codes. Finally, we compare stabilizer and
subsystem codes with respect to the required number of syndrome qudits. 
\end{abstract}

\section{INTRODUCTION}\label{sec:intro}
Quantum error-correcting codes are seen as being indispensable for
building a quantum computer. There are three predominant approaches to
quantum error-correction: stabilizer codes, noiseless systems, and
decoherence free subspaces. Recent advances in the theory of quantum
error-correction have shown that all these apparently disparate
approaches are actually the same. This unification goes under the name
of operator quantum error-correction codes
\cite{kribs05,kribs05b,kribs05c,poulin05,knill06,bacon06}, though we
will prefer to use the shorter and more descriptive term
\textsl{subsystem codes}.  Subsystem codes provide a common platform
for comparing the various different types of quantum codes and make it
possible to treat active and passive quantum error-correction within the same
framework. Apart from the fact that subsystem codes give us more
control over the degree of passive error-correction, there have been
claims that subsystem codes can make quantum error-correction more
robust and practical. For example, it has been claimed that subsystem
codes make it possible to derive simpler error recovery schemes in
comparison to stabilizer codes. Furthermore, it is conjectured that
certain subsystem codes are self-correcting~\cite{bacon06}. 

Subsystem codes are relatively new and promise to be a fruitful area
for quantum error-correction. Until now, there are few concrete
examples of such codes and even fewer systematic code
constructions. Little is known about the parameters of subsystem
codes, so it is difficult to judge the performance of such codes. 

In a recent work \cite{pre7}, we derived a character-theoretic
framework for the construction of subsystem codes. We were able to
show that there exists a correspondence between the subsystem codes
over $\F_q$ and classical additive codes over $\F_q$ and
$\F_{q^2}$. In this paper, we investigate basic properties of
subsystem codes, establish further connections to classical codes, and
derive bounds on the parameters of subsystem codes.  We report first
results on a fair comparison between stabilizer codes
and subsystem codes.
\smallskip

The paper is structured as follows. After a brief introduction to
subsystem codes in Section~\ref{sec:classical}, we recall some results
about the relations between subsystem codes and classical codes. Then
we give some simple constructions of subsystem codes which parallel
the common constructions of stabilizer codes. In
Section~\ref{sec:lBounds}, we give a nonconstructive proof of the
existence of subsystem codes. In Section~\ref{sec:uBounds}, we derive
linear programming upper bounds on the parameters of subsystem
codes. For pure subsystem codes (to be defined later) we can also
derive analytical upper bounds which resemble the quantum Singleton
and Hamming bounds. Armed with these results on bounds we make a
rigorous comparison of stabilizer codes and subsystem codes, that
makes precise when subsystem codes can do better.

{\em Notation:} We assume that $q$ is the power of a prime $p$
and $\F_q$ denotes a finite field with $q$ elements. 
By qudit we mean a $q$-ary quantum bit.  The symplectic
weight of an element $w=(x_1,\ldots,x_n,y_1,\ldots,y_n)$ in
$\F_q^{2n}$ is defined as $\swt(w)=|\{(x_i,y_i)\neq (0,0)\mid 1\leq
i\leq n \} |$. The trace-symplectic product of two elements
$u=(a|b),v=(a'|b')$ in $\F_q^{2n}$ is defined as $\langle u|v
\rangle_s = \tr_{q/p}(a'\cdot b-a\cdot b')$, where $x\cdot y$ is the
usual euclidean inner product. The trace-symplectic dual of a code
$C\subseteq \F_q^{2n}$ is defined as $C^\sdual=\{ v\in \F_q^{2n}\mid
\langle v|w \rangle_s =0 \mbox{ for all } w\in C\}$. For vectors $x,y$
in $\F_{q^2}^n$, we define the hermitian inner product $\langle
x|y\rangle_h =\sum_{i=1}^nx_i^qy_i$ and the hermitian dual of
$C\subseteq \F_{q^2}^n$ as $C^\hdual= \{x\in \F_{q^2}^n\mid \langle
x|y \rangle_h=0 \mbox{ for all } y\in C \}$.  The trace alternating form of two vectors $u,w$
in $\F_{q^2}^n$ is defined as $\langle u|v\rangle_a=\tr_{q/p}[(\langle u|v\rangle_h - \langle v|u\rangle_h)/(\beta^{2}-\beta^{2q})]$, where $\{\beta,\beta^q\}$ is a 
normal basis of $\F_{q^2}$ over $\F_q$. 
If $C\subseteq \F_{q^2}^n$,
then the trace alternating dual of $C$ is defined as $C^\adual =\{x\in \F_{q^2}^n\mid \langle x|y\rangle_a =0\mbox{ for all } y\in C\}$.

\section{SUBSYSTEM CODES AND CLASSICAL CODES} \label{sec:classical}
Let $\mathcal{H}=\C^q\otimes \C^q\otimes \cdots \otimes \C^q
=\C^{q^n}$.  An orthonormal basis for $\C^{q^n}$ is $B=\{\ket{x}\mid x\in \F_q^n \}$.
The vector $\ket{x} =\ket{x_1}\otimes\ket{x_2}\otimes\cdots\otimes \ket{x_n}$.
The elements of $\mathcal{H}$ are of the form 
$$ 
v=\sum_{x\in \F_q^n} v_x \ket{x} \mbox{ where $v_x \in \C$ and }
\sum_{x\in\F_q^n}|v_x|^2 = 1.
$$
We define the following unitary operators on $\C^q$ 
$$ X_a\ket{x}=\ket{x+a} \mbox{ and } Z_b\ket{x} = \omega^{\tr_{q/p}(bx)}\ket{x},$$
where $\omega=e^{j2\pi/p}$. The set of errors $\mathcal{E}= \{ X_a Z_b\mid a,b\in \F_q \}$
form a basis for errors on a single qudit. 
Every error on a single qudit can be expressed as linear combination of the elements
in $\mathcal{E}$. If we assume that the errors are independent on each qudit, we need
 only consider the error group $E=\{\omega^c e_1\otimes e_2 \otimes \cdots \otimes e_n \mid c\in \F_p, e_i\in \mathcal{E}\}$, where
each of the $e_i$ is a single qudit error. The weight of an error is
the number of qudits that are in error. For further details on the
error model and the actual structure of the error group we refer the
reader to \cite{pre3}.  

A quantum error-correcting code $Q$ is a subspace in
$\mathcal{H}=\C^{q^n}$ such that $\mathcal{H}=Q\oplus Q^\perp$, where
$Q^\perp$ is the orthogonal complement of $Q$.  In a subsystem code,
the subspace $Q$ further decomposes into a tensor product of two vector space 
$A$ and $B$, that is,
$$ Q = A\otimes B.$$ The vectors spaces $A$ and $B$ are respectively
called the subsystem and the co-subsystem of the code $Q$.  The
information to be protected is stored in the subsystem $A$, whence the
name subsystem code. 

If $\dim A=K$, $\dim B=R$ and $Q$ is able to detect all errors in $E$
of weight less than $d$ on subsystem $A$, then we say that $Q$ is an
$((n,K,R,d))_q$ subsystem code.  We call $d$ the minimum distance of
the subsystem $A$ or, by slight abuse of language, the minimum
distance of the subsystem code $Q$ (when the tensor decomposition
$Q=A\otimes B$ is understood from the context). We write
$[[n,k,r,d]]_q$ for an $((n,q^k,q^r,d))_q$ subsystem code.  Sometimes
we will say that an $[[n,k,r,d]]_q$ subsystem code has $r$ virtual
gauge qudits, which is simply another way of saying that the dimension
of the co-subsystem is $q^r$; it should be stressed that the gauge
qudits typically do not correspond to physical qudits.

\subsection{Subsystem Codes From Classical Codes}
We recall the following results from \cite{pre7} which relate quantum
subsystem codes to classical codes.

\begin{theorem}\label{th:oqecfq}
Let $X$ be a classical additive subcode of\/ $\F_q^{2n}$ such that $X\neq
\{0\}$ and let $Y$ denote its subcode $Y=X\cap X^\sdual$. If $x=|X|$ and
$y=|Y|$, then there exists subsystem code $C=
A\otimes B$ such that
\begin{compactenum}[i)]
\item $\dim A = q^n/(xy)^{1/2}$,
\item $\dim B = (x/y)^{1/2}$.
\end{compactenum}
The minimum distance of subsystem $A$ is given by
$d=\swt((X+X^\sdual)-X)=\swt(Y^\sdual-X)$. Thus, the subsystem $A$ can detect
all errors in $E$ of weight less than $d$, and can correct all errors in $E$ of
weight $\le \lfloor (d-1)/2\rfloor$.
\end{theorem}
\begin{proof}
See \cite[Theorem~5]{pre7}.
\end{proof}
\smallskip

\begin{remark}
Recall that $|X^\sdual|=q^{2n}/|X|$. Therefore, the dimension of the
subsystem $A$ can also be calculated as $\dim A=(|X^\sdual|/|Y|)^{1/2}$. 
\end{remark}
\smallskip

It is also possible to construct subsystem codes via codes over $\F_{q^2}$ using
$\langle\,\cdot\,|\,\cdot\,\rangle_a$,
the trace alternating form \cite{pre3} which gives us the following theorem. 
The proof can be found in \cite[Theorem~6]{pre7}. 
\begin{theorem}\label{th:oqecfq2}
Let $X$ be a classical additive subcode of\/ $\F_{q^2}^{n}$ such that $X\neq
\{0\}$ and let $Y$ denote its subcode $Y=X\cap X^\adual$. If $x=|X|$ and
$y=|Y|$, then there exists subsystem code $C=
A\otimes B$ such that
\begin{compactenum}[i)]
\item $\dim A = q^n/(xy)^{1/2}$,
\item $\dim B = (x/y)^{1/2}$.
\end{compactenum}
The minimum distance of subsystem $A$ is given by
$$d=\wt((X+X^\adual)-X)=\wt(Y^\adual-X),$$
where $\wt$ denotes the Hamming weight.  Thus, the subsystem $A$ can detect all
errors in $E$ of Hamming weight less than $d$, and can correct all errors in
$E$ of Hamming weight $\lfloor (d-1)/2\rfloor$ or less.
\end{theorem}
\begin{proof}
This follows from Theorem~\ref{th:oqecfq} and the fact that there
exists a weight-preserving isometric isomorphism from
$(\F_{q}^{2n},\langle\,\cdot\,|\,\cdot\,\rangle_s)$ and
$(\F_{q^2}^{n},\langle\,\cdot\,|\,\cdot\,\rangle_a)$, see~\cite{pre3}.
\end{proof}

Theorem~\ref{th:oqecfq2} has the advantage that the weights of the
codes over $\F_{q^2}$ is measured using the usual Hamming distance.

We are now going to derive some particularly important special cases
of the above two theorems as a consequence. Before stating these
results, we recall the following simple fact.

\begin{lemma}\label{th:dirsumdual}
Let $C_1$ and $C_2$ be two $\F_q$-linear codes of length $n$. The product code 
$C_1\times C_2 = \{ (a|b) \,|\, a\in C_1,
b\in C_2\}$ has length $2n$ and its trace-symplectic dual is given by
$$ (C_1\times C_2)^\sdual = C_2^\perp\times C_1^\perp.$$
\end{lemma}
\begin{proof}
If $(a|b)\in C_1\times C_2$ and $(a'|b')\in C_2^\perp\times C_1$, then 
$ \tr_{q/p}(b\cdot a' - b'\cdot a)= 0;$
hence, $C_2^\perp\times C_1^\perp\subseteq (C_1\times C_2)^\sdual$. 
Comparing dimensions shows that equality must hold. 
\end{proof}
\smallskip

The first consequence uses the euclidean inner product, that is, the
usual dot inner product on $\F_q^n$ to construct subsystem codes. In
the special case of stabilizer codes, this yields the well-known CSS
construction (for instance, see \cite[Theorem~9]{calderbank98}).

\begin{corollary}[Euclidean Construction]\label{th:cssoqec}
Let $C_i \subseteq \F_q^n$, be $[n,k_i]_q$ linear codes where $i\in \{1,2\}$. Then 
there exists an $[[n,k,r,d]]_q$ subsystem code with
\begin{compactitem}
\item $k=n-(k_1+k_2+k')/2$, 
\item $r=(k_1+k_2-k')/2$, and 
\item 
$d=\min \{ \wt((C_1^\perp\cap C_2)^\perp\setminus C_1), 
\wt((C_2^\perp\cap C_1)^\perp\setminus C_2) \}$,
\end{compactitem}
where $k'= \dim_{\F_q}(C_1\cap C_2^\perp)\times (C_1^\perp\cap C_2)$. 
\end{corollary} 
\smallskip
\begin{proof}
Let $C=C_1\times C_2$, then by Lemma~\ref{th:dirsumdual}, 
$C^\sdual=C_2^\perp\times C_1^\perp$, and $D=C\cap C^\sdual = (C_1\cap
C_2^\perp)\times (C_2\cap C_1^\perp)$. Again by
Lemma~\ref{th:dirsumdual}, $D^\sdual = (C_2\cap C_1^\perp)^\perp
\times (C_1\cap C_2^\perp)^\perp$.  Let $\dim_{\F_q} D= k'$. Then
$|C||D|=q^{k_1+k_2+k'}$ and $|C|/|D|=q^{k_1+k_2-k'}$.  By
Theorem~\ref{th:oqecfq}, the code $C$ defines an
$[[n,n-(k_1+k_2+k')/2,(k_1+k_2-k')/2,d]]_q$ subsystem code. The
distance of the code is given by $$
\begin{array}{lcl}
d&=&\swt(D^\sdual\setminus C)\\ 
&=&
\swt((C_2\cap C_1^\perp)^\perp \times (C_1\cap C_2^\perp)^\perp
\setminus (C_1\times C_2)).
\end{array}
$$ 
The latter expression can be simplified to 
$$d=\min\{
\wt((C_2\cap C_1^\perp)^\perp\setminus C_1), \wt((C_1\cap
C_2^\perp)^\perp\setminus C_2)\},$$
which proves the claim.  
\end{proof}

Setting $C_2=C_1$ in the previous construction simplifies the computation
of the code parameters. Then we have an $[[n,n-k-k', k-k',
\wt((C_1\cap C_1^\perp)^\perp\setminus C_1)]]_q$ code, where
$k'=\dim_{\F_q} C_1\cap C_1^\perp$.  Therefore, any family of
classical codes where the dimension of $C_1\cap C_1^\perp$ and the
minimum distance of the dual of $C_1\cap C_1^\perp$ is known, will
provide us with a family of subsystem codes. The codes that arise when
$C_1=C_2$ will also arise as a special case of the next construction. 
\smallskip
\begin{corollary}[Hermitian Construction]\label{th:oqecHerm}
Let $C \subseteq \F_{q^2}^n$ be an $\F_{q^2}$-linear $[n,k,d]_{q^2}$
code such that $D=C\cap C^\hdual$ is of dimension $k'=\dim_{\F_{q^2}}
D$. Then there exists an
$$[[n,n-k-k',k-k',\wt(D^\hdual \setminus C)]]_q$$ subsystem code.
\end{corollary} 
\begin{proof}
If $C$ is linear, then $C^\adual=C^\hdual$ by \cite[Lemma~18]{pre7}.
It follows that $q^n/\sqrt{|D||C|}=q^{n-k'-k}$ and
$\sqrt{|C|/|D|}=q^{k-k'}$.  Let $d=\wt(D^\hdual\setminus C)$.  Then, by
Theorem~\ref{th:oqecfq2}, there exists an
$[[n,n-k-k',k-k',d]]_q$ subsystem code.
\end{proof}
The subsystem codes can be easily constructed with the help of a computer
algebra system. The following example gives some 
subsystem codes constructed using MAGMA \cite{magma}. 
\begin{example}[BCH Subsystem Codes]\label{ex:herm}
The binary subsystem codes in Table~\ref{bchtable} were derived from 
BCH codes over $\F_{4}$ via Corollary~\ref{th:oqecHerm}. 
\begin{table}[ht]
\caption{BCH Subsystem Codes} \label{bchtable}
\begin{center}
\begin{tabular}{|c|c|c|}
\hline
\text{Subsystem Code} &  \text{Parent} & \text {Designed}  \\
 &  \text{BCH Code} & \text{ distance }  \\

\hline $[[15,1,2,5]]_2$ & $[15,8,6]_{2^2}$ &6 \\
{} $[[15,5,2,3]]_2$&$[15,6,7]_{2^2}$&7\\
 \hline
$[[17,8,1,4]]_2 $&$ [17,5,9]_{2^2}$ &4 \\
\hline
 $[[21,6,3,3]]_2$&$ [21,9,7]]_{2^2}$&6\\{}
 $[[21 ,7 ,2 ,3 ]]_2$& $ [21 ,8 ,9 ]_{2^2}$&8\\
\hline $[[31,10,1,5]]_2$&$[31,11,11]_{2^2} $&8\\{}
$[[31 ,20,1 ,3 ]]_2$&$ [31 ,6 ,15]_{2^2}$&12\\
 \hline
\end{tabular}
\end{center}
\end{table}
\end{example}

Codes constructed with the help of Corollaries~\ref{th:cssoqec} and
\ref{th:oqecHerm} will lead to $\F_q$-linear  and 
$\F_{q^2}$-linear subsystem codes respectively.  Though in some 
cases Corollary~\ref{th:cssoqec} can lead to $\F_{q^2}$-linear codes. 
So when we refer to a subsystem code as being $\F_q$-linear, it could be also 
$\F_{q^2}$-linear. In this
paper, we will call a subsystem code that can be constructed with the
help of Theorems~\ref{th:oqecfq} and~\ref{th:oqecfq2} and their
corollaries, a \textit{Clifford subsystem code}. Ten years ago, Knill
suggested a generalization of stabilizer codes that became known as
Clifford codes (because their construction uses a part of
representation theory known as Clifford theory). Recently, we realized
that a special case of Knill's construction leads to a very natural
construction of subsystem codes. Clifford theory is the natural tool
in the construction of these subsystem codes, whence the name.

\section{LOWER BOUNDS ON SUBSYSTEM CODES} \label{sec:lBounds}
In this section we give a simple nonconstructive proof for the
existence of subsystem codes.  The proof is based on a counting
argument similar to the quantum Gilbert-Varshamov bound for stabilizer
codes \cite{pre3}. We will need the following simple fact. 

\begin{lemma}\label{l:addcode}
Let $\F_q$ be a finite field of characteristic $p$. Let $r$ and $s$ be
nonnegative integers such that $p^{r+2s}\leq q^{2n}$.  Then there exists an
additive subcode $X$ of\/ $\F_q^{2n}$ such that $|X|=p^{r+2s}$ and
$|X\cap X^\sdual|=p^r$.
\end{lemma}
\begin{proof}
Let $m$ denote the integer such that $q=p^m$. We may regard
$\F_q^{2n}$ as an $2nm$-dimensional vector space over $\F_p$.  Then
$\scal{\cdot}{\cdot}$ is a nondegenerate skew-symmetric bilinear form
on this vector space.  Therefore, there exists a direct sum
decomposition of $\F_q^{2n}\cong \F_p^{2nm}=V_1\oplus\cdots \oplus
V_{nm}$, where $V_k$ is a 2-dimensional subspace with basis
$\{x_k,z_k\}$ such that $\scal{x_k}{x_\ell}=0=\scal{z_k}{z_\ell}$ for
$1\le k,\ell\le nm$, $\scal{x_k}{z_k}\neq 0$, and
$\scal{x_k}{z_\ell}=0$ if $k\neq \ell$.  Then $X=\langle
z_1,\dots,z_r, x_{r+1},z_{r+1},\dots, x_{r+s},z_{r+s}\rangle$ is a
code with the desired properties.
\end{proof}

\begin{theorem}\label{th:gvoqec}
Let $\F_q$ be a finite field of characteristic $p$.  If $K$ and $R$
are powers of $p$ such that $1<KR\le q^n$ and $d$ is a positive
integer such that
$$
\sum_{j=1}^{d-1} \binom{n}{j}(q^{2}-1)^j (q^nKR-q^nR/K)<(p-1)(q^{2n}-1)$$
holds, then an
$((n,K,R,\ge d))_q$ subsystem code exists. 
\end{theorem}
\begin{proof}
By Lemma~\ref{l:addcode}, there exists an additive subcode $X$ of
$\F_q^{2n}$ such that $x=|X|=q^nR/K$ and $y=|X\cap
X^\sdual|=q^n/(KR)$; the resulting subsystem code has a subsystem of
dimension $q^n/(xy)^{1/2}=K$ and a co-subsystem of dimension
$(x/y)^{1/2}=R$. Therefore, the multiset $\mathcal{X}$ given by
$$ \mathcal{X} = \left\{ (X+X^\sdual)-X\,\bigg|\, 
\begin{array}{l}
X \text{ is an additive subcode of } \\
\F_q^{2n} \text{ such that } 
|X|=q^nR/K \\
\text{and } |X\cap X^\sdual|=q^n/(KR)
\end{array}
\right\}$$ is not empty. 

Thus, an element of $\mathcal{X}$ corresponds to a subsystem code
$C=A\otimes B$ with $\dim A=K$ and $\dim B=R$.  The set difference
$(X+X^\sdual)-X$ contains only nonzero vectors of $\F_q^{2n}$. We
claim that all nonzero vector in $\F_q^{2n}$ appear in the same number
of sets in $\mathcal{X}$. Indeed, the symplectic group
$\textup{Sp}(2n,\F_q)$ acts transitively on the set
$\F_{q}^{2n}\setminus \{ 0\}$, see~\cite[Proposition~3.2]{grove01},
which means that for any nonzero vectors $u$ and $v$ in $\F_q^{2n}$
there exists $\tau\in \textup{Sp}(2n,\F_q)$ such that $v=\tau
u$. Therefore, $u$ is contained in $(X+X^\sdual)-X$ if and only if $v$
is contained in the element $(\tau X+(\tau X)^\sdual)-\tau X$ of
$\mathcal{X}$.

Since $|(X+X^\sdual)-X|=q^nKR-q^nR/K$, we can conclude that any
nonzero vector of $\F_q^{2n}$ occurs in
$|\mathcal{X}|(q^nKR-q^nR/K)/(q^{2n}-1)$ elements of $\mathcal{X}$.
Furthermore, a nonzero vector and its $\F_p^\times$-multiples are
contained in the exact same sets of $\mathcal{X}$. Therefore, if we
delete all sets from $\mathcal{X}$ that contain a nonzero vector with
symplectic weight less than $d$, then we remove at most
$$
\frac{\sum_{j=1}^{d-1} \binom{n}{j}(q^{2}-1)^j}{p-1}
|\mathcal{X}|\frac{(q^nKR-q^nR/K)}{q^{2n}-1}$$ sets from $\mathcal{X}$. 
By assumption, this
number is less than $|\mathcal{X}|$; hence, there exists an 
$((n,K,R,\ge d))_q$ subsystem code. 
\end{proof}
The lower bound has important implications for comparing stabilizer codes
with subsystem codes as we shall see in Section~\ref{sec:stabVsoqec}. 
Further, we obtain the following lower bound for stabilizer codes as a
simple corollary, when $R=1$ (see also \cite{pre3}).
\begin{corollary}[GV Bound for Stabilizer Codes]\label{th:gvstab}
Let $\F_q$ be a finite field of characteristic $p$ and $1<K\leq q^n$ a power of $p$. If $$
\sum_{j=1}^{d-1} \binom{n}{j}(q^{2}-1)^j <(p-1)\frac{(q^{2n}-1)}{(q^nK-q^n/K)}$$
holds,  then an $((n,K,\ge d))_q$ stabilizer code exists.
\end{corollary}
A stronger result showing the existing of linear stabilizer codes was shown in 
\cite[Lemma~31]{pre3}.
%
\section{UPPER BOUNDS FOR SUBSYSTEM CODES} \label{sec:uBounds}
We want to investigate some limitations on subsystem codes that can be
constructed with the help of Theorem~\ref{th:oqecfq} (or,
equivalently, Theorem~\ref{th:oqecfq2}).  To that end, we will
investigate some upper bounds on the parameters of subsystem codes.


\subsection{Linear Programming Bounds}
\begin{theorem}\label{th:lp}
If an $((n,K,R,d))_q$ Clifford subsystem code with
$K>1$ exists,
then there exists a solution to the optimization
problem: maximize $\sum_{j=1}^{d-1} A_j$ subject to the constraints
\begin{enumerate}
\item $A_0=B_0=1$ and $0\le B_j \le A_j$ for all $1\le j\le n$;
\item  $\ds\sum_{j=0}^n A_j = q^{n}R/K$; \quad $\ds\sum_{j=0}^n B_j = q^{n}/KR$;
\item $A_j^\sdual = \ds\frac{K}{q^{n}R} \sum_{r=0}^n K_j(r)A_r$ holds for all $j$ in the 
range $0\le j \le n $;
\item $B_j^\sdual = \ds\frac{KR}{q^{n}} \sum_{r=0}^n K_j(r)B_r$ holds for all $j$ in the 
range $0\le j \le n $;
\item $A_j=B_j^\sdual$ for all $j$ in $0\le j<d$ and $A_j\le B_j^\sdual$ for all $d\le j\le n$;
\item $B_j=A_j^\sdual$ for all $j$ in $0\le j<d$ and $B_j\le A_j^\sdual$ for all $d\le j\le n$;
\item $(p-1)$ divides $A_j$, $B_j$, $A_j^\sdual$, and $B_j^\sdual$ for
all $j$ in the range $1\le j\le n$;
\end{enumerate}
where the coefficients $A_j$ and $B_j$ assume only integer values, and
$K_j(r)$ denotes the Krawtchouk polynomial
$$ K_j(r) = \sum_{s=0}^j (-1)^s (q^2-1)^{j-s}\binom{r}{s}\binom{n-r}{j-s}.$$ 
\end{theorem}
\begin{proof}
If an $((n,K,R,d))_q$ subsystem code exists,
then the weight distribution $A_j$ of the associated additive code~$X$
and the weight distribution $B_j$ of its subcode $Y =X\cap X^\sdual$
obviously satisfy~1).  By Theorem~\ref{th:oqecfq}, we have
$K=q^n/\sqrt{|X||Y|}$ and $R=\sqrt{|X|/|Y|}$, which implies $|X|=\sum
A_j = q^nR/K$ and $|Y|=\sum B_j =q^n/KR$, proving~2). Conditions~3)
and~4) follow from the MacWilliams relation for symplectic weight
distribution, see \cite[Theorem~23]{pre3}. As $X$ is an $\F_p$-linear
code, for each nonzero codeword $c$ in $X$, $\alpha c$ is again in $X$
for all $\alpha$ in $\F_p^\times$; thus, condition~7) must hold.
Since the quantum code has minimum distance $d$, all vectors of
symplectic weight less than $d$ in $Y^\sdual$ must be in $X$, since
$Y^\sdual-X$ has minimum distance $d$; this implies~5). Similarly, all
vectors in $X^\sdual\subseteq X+X^\sdual$ of symplectic weight less
than $d$ must be contained in $X$, since $(X+X^\sdual)-X$ has minimum
distance $d$; this implies~6).
\end{proof}
\smallskip

We can use the previous theorem to derive bounds on the dimension 
of the co-subsystem. If the optimization problem is not solvable, then
we can immediately conclude that a code with the corresponding
parameter settings cannot exist.  

Perhaps one of the most striking features of subsystem codes is the
potential reduction of syndrome measurements. Recall that an
$\F_q$-linear $[[n,k,d]]_q$ stabilizer code requires $n-k$ syndrome
measurements. On the other hand, an $\F_q$-linear $[[n,k,r,d]]_q$
Clifford subsystem code requires just $n-k-r$ syndrome measurements. 

Poulin~\cite{poulin05} asked whether we can have $[[5,1,r>0,3]]_2$
Clifford subsystem code. Of course, such a code would be preferable
over the $[[5,1,3]]_2$ stabilizer code. After an exhaustive computer
search, he concluded that such a subsystem code does not exist. This
result can be obtained very easily with the linear programming
bounds. In fact, our investigations for small lengths revealed that
not only a $[[5,1,r>0,3]]_2$ code does not exist, but neither does any
code with parameters given in the next example.

\begin{example}\label{ex:lp}
Theorem~\ref{th:lp} shows that it is not possible to construct
subsystem codes with $r>0$ and parameters shown in Table~\ref{lptable}.
\begin{table}[ht]
\caption{} \label{lptable}
\begin{center}
\begin{tabular}{|c|c|}
\hline
\text{Field} & \quad \text{Codes}  \\
\hline
& \\[-2ex]
$\F_2$ &$ [[4,2,r,2]]_2$,\; $[[5,1,r,3]]_2$\\
\hline
 & $[[4,2,r,2]]_3$,\; $[[5,1,r,3]]_3$,\;\\
 $\F_3$& $[[9,3,r,4]]_3$,\; $[[9,5,r,3]]_3$,\; \\
 &$[[10,6,r,3]]_3$ \\
\hline
& $[[4,2,r,2]]_4$,\; $[[5,1,r,3]]_4$,\;\\
$\F_4$ & $[[9,3,r,4]]_4$,\; $[[9,5,r,3]]_4$,\;\\
& $[[10,6,r,3]]_4$\;\\
\hline
\end{tabular}
\end{center}
\end{table}
\end{example}
\smallskip

The previous example is motivated by the fact that one can improve
upon Shor's $[[9,1,3]]_2$ quantum stabilizer code by allowing three
additional gauge qubits, that is, there exists a $[[9,1,3,3]]_2$
subsystem code, see~\cite{poulin05}. The practical relevance is that
the $9-1=8$ syndrome measurements that are required for Shor's code
are reduced to $9-1-3=5$ syndrome measurements in the subsystem code. 

Since we allow nonbinary alphabets in this paper, a natural
generalization of Poulin's question is whether one can find an
$[[n,n-2d+2,r,d]]_q$ subsystem code with $r>0$. The above example
shows that such subsystem codes with such parameters do not exist for
certain small lengths and small alphabet sizes.

We will fully answer this question in the subsequent sections. In
the search for an answer to this problem, we were prompted to define the
notion of pure subsystem codes. The notion of purity proved to be
fruitful in deducing this and other results.

\subsection{Pure Subsystem Codes} 
Let $X$ be an additive subcode of $\F_{q}^{2n}$ and $Y=X\cap
X^\sdual$.  By Theorem~\ref{th:oqecfq}, we can obtain an
$((n,K,R,d))_q$ subsystem code $Q$ from $X$
that has minimum distance $d=\swt(Y^\sdual - X)$.  The set difference
involved in the definition of the minimum distance makes it harder to
compute the minimum distance. Therefore, we introduce pure
codes that are easier to analyze. 

We say that the subsystem code $Q$ is \textit{pure to $d'$} if $d'\leq
\swt(X)$.  The code is \textit{exactly pure to $d'$} if it is pure to
$d'$ but not to $d'+1$; then $\swt(X)=d'$. Any subsystem code is always
exactly pure to $d'=\swt(X)$.  We call $Q$ a pure subsystem code if it is pure
to $d'\geq d$; otherwise, we call $Q$ an impure subsystem code. Pure
codes do not require us to compute the minimum distance of the
difference set $Y^\sdual - X$. We can compute the distance of the code
as $d=\swt(Y^\sdual)$, which is comparatively simpler task though it
is also computationally hard.

The purity of codes over $\F_{q^2}$ is defined in a similar way. 

\begin{example}[Reed-Solomon Subsystem Codes]\label{ex:pure}
The nonbinary subsystem codes given in Table~\ref{tab:RS} are all pure and
were derived from primitive narrowsense Reed-Solomon codes over $\F_{q^2}$.
\begin{table}[ht]
\caption{Reed-Solomon Subsystem codes} \label{rstable}
\begin{center}
\begin{tabular}{|c|c|c|}
\hline\hline
\text{Subsystem Codes} &  \text{Parent}  \\
 &  \text{RS Code}  \\
\hline $[[15,1,10,3]]_4$ & $[15 ,12 ,4 ]_{4^2}$  \\{}
 $[[15 ,1  ,8  ,3 ]]_4$ & $[15 ,11 ,5]_{4^2}$ \\{}  $[[15 ,1  ,6  ,3
]]_4$&$[15 ,10 ,6]_{4^2}$\\{} $[[15 ,2  ,5  ,3  ]] _4$&$ [15 ,9  ,7]_{4^2}$\\ \hline
$[[24,1,17,4]]_5$ &$[24,20,5]_{5^2}$
\\{} $[[24,2,10,4]]_5 $&$[24,16,9]_{5^2}$\\{}
$[[24 ,4 ,10,4 ]]_5$ &$[24,15,10]_{5^2}$\\{}$ [[24,16,2,4]]_5$
&$[24,5,20]_{5^2}$\\{}
$[[24,17 ,1,4 ]]_5 $&$[24,4,21]_{5^2}$\\{} $[[24,19,1,3]]_5$ &$[24,3,22]_{5^2}$\\
 \hline
 $[[48 ,1  ,37 ,6  ]]_7$  &$[48 ,42 ,7 ]_{7^2}$\\{}
  $[[48 ,2  ,26 ,6  ]]_7 $ &$[48 ,36 ,13 ]_{7^2}$\\
 \hline
\end{tabular}
\end{center}\label{tab:RS} 
\end{table}
It is curious that the distance of many of these subsystem codes is
equal to $q-1$.  We conjecture that, in general, the distance of a
subsytem code constructed from a Reed-Solomon code over $\F_{q^2}^{q^2-1}$
cannot exceed $q-1$.
\end{example}

\subsection{Upper Bounds for Pure Subsystem Codes}
In this subsection, we establish a number of basic results concerning
pure subsystem codes. The next lemma is a key result that associates
to a pure subsystem code a pure stabilizer code.
\begin{lemma}\label{th:stabcode}
If a pure $((n,K,R,d))_q$ Clifford subsystem code~$Q$ exists, then
there exists a pure $((n,KR,d))_q$ stabilizer code.
\end{lemma}
\begin{proof}
Let $X$ be a classical additive subcode of $\F_q^{2n}$ that defines
$Q$, and let $Y=X\cap X^\sdual$.  Furthermore, Theorem~\ref{th:oqecfq} implies 
that $KR=q^n/|Y|$. Since $Y\subseteq Y^\sdual$, there exists an
$((n,q^n/|Y|,d'))_q$ stabilizer code with minimum distance
$d'=\wt(Y^\sdual - Y)$. The purity of $Q$ implies that 
$\swt(Y^\sdual - X) = \swt(Y^\sdual)=d$. As $Y\subseteq X$, it follows
that $d'=\swt(Y^\sdual - Y)=\swt(Y^\sdual)=d$; hence, there
exists a pure $((n,KR,d))_q$ stabilizer code.
\end{proof}

As a consequence of the preceding lemma, it is straightforward to
obtain the following bounds on pure subsystem codes.
\begin{theorem}\label{th:pureBound} 
Any pure $((n,K,R,d))_q$ Clifford subsystem code 
satisfies 
$KR\leq q^{n-2d+2}$.
\end{theorem}
\begin{proof}
By Lemma~\ref{th:stabcode}, there exists a pure $((n,KR,d))_q$
stabilizer code. By the quantum Singleton bound, we have
$KR\leq q^{n-2d+2}$.
\end{proof}
\begin{corollary}
A pure $[[n,k,r,d]]_q$ Clifford subsystem code satisfies 
$ k+ r\leq n-2d+2$.
\end{corollary}
\begin{example}[Optimal Subsystem Codes]\label{ex:mds}
All the following codes constructed from Reed-Solomon codes over $\F_{q^2}$ are
pure and meet the bound in Theorem~\ref{th:pureBound}. These codes are in that 
sense optimal subsystem codes. 

\begin{table}[ht]
\caption{Optimal Pure Subsystem Codes} \label{opttable}
\begin{center}
\begin{tabular}{|c|c|c|}
\hline
\text{Subsystem Codes} &  \text{Parent}  \\
 &  \text{Code (RS Code)}  \\
\hline $[[15,1,10,3]]_4$ & $[15 ,12 ,4 ]_{4^2}$  \\{}
$[[15,9,2,3]]_4$&$[15,4,12]_{4^2}$\\{} $[[15,10,1,3]]_4$&$[15,3,13]_{4^2}$\\
  \hline $[[24,1,17,4]]_5$
&$[24,20,5]_{5^2}$
\\{}
$ [[24,16,2,4]]_5$ &$[24,5,20]_{5^2}$\\{}
$[[24,17 ,1,4 ]]_5 $&$[24,4,21]_{5^2}$\\{}
 $[[24,19,1,3]]_5$ &$[24,3,22]_{5^2}$\\
 \hline
 $[[48 ,1  ,37 ,6  ]]_7$  &$[48 ,42 ,7 ]_{7^2}$\\
 \hline
\end{tabular}
\end{center}
\end{table}
\end{example}

We can also show that the pure subsystem codes obey a quantum
Hamming bound like the stabilizer codes.  We skip the proof as it is
along the same lines as Theorem~\ref{th:pureBound}.
\begin{lemma}
A pure $((n,K,R,d))_q$ Clifford subsystem code satisfies
$$ \sum_{j=0}^{\lfloor\frac{d-1}{2} \rfloor}\binom{n}{j}(q^2-1)^j \leq
q^n/KR.$$
\end{lemma}

\section{SUBSYSTEM CODE CONSTRUCTIONS}\label{sec:const}
In this section, we give new constructions for pure subsystem
codes. We begin with a proof of the simple, yet surprising,
observation that one can always exchange information qudits and gauge
qudits in the case of pure subsystem codes.

\begin{lemma}
If there exists a pure $((n,K,R,d))_q$ Clifford subsystem code, then
there also exists an $((n,R,K, \mbox{$\ge$}\,d))_q$ Clifford
subsystem code that is pure to $d$.
\end{lemma}
\begin{proof}
By Theorem~\ref{th:oqecfq2}, there exist classical codes $D\subseteq
C\subseteq \F_{q^2}^n$ with the parameters $(n,q^nR/K)_{q^2}$ and
$(n,q^n/KR)_{q^2}$.  Furthermore, since the subsystem code is pure, we have 
$\wt(D^\adual\setminus C) = \wt( D^\adual)= d$.

Let us interchange the roles of $C$ and $C^\adual$, that is, now we
construct a subsystem code from $C^\adual$. The parameters of the
resulting subsystem code are given by
$$((n, \sqrt{|D^\adual|/|C^\adual|},\sqrt{|C^\adual|/|D|},\wt(D^\adual\setminus C^\adual) ))_q.$$ 
We note that 
\begin{compactitem}
\item  $\sqrt{|D^\adual|/|C^\adual|} =\sqrt{|C|/|D|} =R$ and 
\item $\sqrt{|C^\adual|/|D|}= \sqrt{|D^\adual|/|C|}=K$. 
\end{compactitem}
The minimum distance $d'$ of the resulting code satisfies $d' =
\wt(D^\adual\setminus C^\adual) \geq \wt( D^\adual) = d$; the claim
about the purity follows from the fact that $\wt(D^\adual)=d$.
\end{proof}

Before proving our next result, we need the following fact from linear
algebra.

\begin{lemma}\label{l:hyperbolic_basis}
Let $\F_q$ be a finite field of characteristic $p$.  
Let $C$ denote an additive subcode of $\F_q^{2n}$.
There exists an $\F_p$-basis $B$ generating the code $C$ that is of the form
$$ B=\{z_1,x_1;\dots;z_r,x_r;z_{r+1},\dots,z_{r+j}\}$$ where
$\scal{x_k}{x_\ell}=0=\scal{z_k}{z_\ell}$ and
$\scal{x_k}{z_\ell}=\delta_{k,\ell}$.  In particular, $D=C\cap
C^\sdual=\langle z_{r+1},\dots,z_{r+s}\rangle$.  It is possible to choose $B$
such that it contains a vector $z_k$ of minimum weight $\swt(C)$.
\end{lemma}
\begin{proof}
Choose a basis $\{z_1,\dots,z_{r+j}\}$ of a maximal isotropic subspace
$C_0$ of $C$. If $C_0\neq C$, then we can choose a codeword $x_1$ in
$C$ that is orthogonal to all of the $z_k$ except one, say $z_1$
(renumbering if necessary).  By multiplying with a scalar in
$\F_p^\times$, we may assume that $\scal{z_1}{x_1}=1$.  If $\langle
C_0,x_1\rangle\neq C$, then one can repeat the process a finite number
of times by choosing an $x_k$ that is orthogonal to $\{x_1,\dots,x_{k-1}\}$ 
until a basis of the desired form is found.
\end{proof}

A subset $\{z_k,x_k\}$ of $C$ with $\scal{z_k}{x_k}=1$ is called a
hyperbolic pair. Thus, in the proof of the previous lemma, one chooses
in each step a hyperbolic pair that is orthogonal to the previously
chosen hyperbolic pairs.

\begin{theorem}[`Rain on your Parade Theorem']\label{th:gaugeReduction}
Let $\F_q$ be a finite field of characteristic $p$. An
$((n,K,R>1,d))_q$ Clifford subsystem code $Q$ implies the existence
of an $((n,K,R/p,d))_q$ Clifford subsystem code $Q_s$.  If $Q$ is
exactly pure to $d'$, then the subsystem code $Q_s$ can be chosen 
such that it is exactly pure to $d'$ as
well.
\end{theorem}
\begin{proof}
By Theorem~\ref{th:oqecfq}, there exists an additive code $C\le
\F_{q}^{2n}$ with subcode $D=C\cap C^\sdual$ such that
$K=q^n/(|D||C|)^{1/2}$, $R=(|C|/|D|)^{1/2}$,
$d=\swt(D^\sdual\setminus C)$, and $d'=\swt(C)$. By
Lemma~\ref{l:hyperbolic_basis}, one can find a $\F_p$-basis $B$ of the form
$ B=\{z_1,x_1;\dots;z_r,x_r;z_{r+1},\dots,z_{r+j}\}$ such that
$\scal{x_k}{x_\ell}=0=\scal{z_k}{z_\ell}$ and
$\scal{x_k}{z_\ell}=\delta_{k,\ell}$. Notice that $D=C\cap
C^\sdual=\langle z_{r+1},\dots,z_{r+j}\rangle$ by Lemma~\ref{l:hyperbolic_basis}.

Let $C_s$ be the additive subcode of $C$ given by $C_s =
\Span_{\F_p}(B\setminus\{x_r\}).$ Then $D_s=C_s\cap C_s^\sdual=\langle
z_{r},\dots,z_{r+j}\rangle$.  It follows that
$|C_s|=|C|/p$ and $|D_s|=p|D|$. Therefore, $C_s$ defines a subsystem
code $Q_s=A_s\otimes B_s$ such that 
$\dim A_s = q^n/(|C_s||D_s|)^{1/2}=K$
and $\dim B_s = (|C_s|/|D_s|)^{1/2} =R/p$. 

Since $D_s^\sdual \subset D^\sdual$, any minimum weight codeword $c\in
D_s^\sdual\setminus C_s$ must be either in $D^\sdual \setminus C$ or
$C$. If it is in $D^\sdual \setminus C$, then $\swt(c)\geq d$. If it
is in $C$, then it is a linear combination of elements in $B\setminus
\{x_r\}$, since $x_r \not\in D_s^\sdual$. This implies that $c$ is contained 
in $C_s$, contradicting our assumption that $c$ is in
$D_s^\sdual \setminus C_s$.  Therefore, $\swt(D_s^\sdual \setminus C_s
)\geq d$ and we can conclude that $Q_s$ has minimum distance $\ge d$. 

For the purity statement, recall that $D\subset D_s\subseteq
C_s\subset C$. The subsystem code $Q$ is exactly pure to $d'=\swt(C)$.  If
$\swt(D)=d'$, then $\swt(C_s)=d'$; otherwise, $\swt(C\setminus D)=d'$
and we can choose $z_{r+1}$ such that $\swt(z_{r+1})=d'$. Then the
subsystem code $Q_s$ is exactly pure to $\swt(C_s)=d'$.
\end{proof}

\smallskip

\begin{corollary}\label{th:oqec2stab}
An $((n,K,R,d))_q$ Clifford subsystem code that is exactly pure to $d'$
implies the existence of an $((n,K,\geq d))_q$ stabilizer code that is
(exactly) pure to $d'$.
\end{corollary}
\begin{proof}
The corollary follows by repeatedly applying
Theorem~\ref{th:gaugeReduction} to the $((n,K,R,d))_q$ code and the derived
code 
until the dimension of the gauge subsystem is reduced to one. 
\end{proof}

We know that the MDS stabilizer codes arise from classical MDS
codes. In fact, the stabilizer code is MDS if and only if the
associated classical code is MDS. We can therefore hope that good
subsystem codes can be obtained from classical MDS codes.  We show
that the resulting subsystem codes must be pure.

\begin{lemma}
If an $((n,K>1,R>1,d))_q$ subsystem code is constructed from an MDS code, then the resulting code is pure.
\end{lemma}
\begin{proof}
Assume that $C\subseteq \F_{q^2}^n $ is an $[n,k,n-k+1]_{q^2}$ code. If 
$C^\adual \subseteq C$, then $K=1$ contrary to our assumption.
So assume that $C^\adual \not\subseteq C$. Let $k>n-k$.
Then $D=C\cap C^\adual$ must be smaller than $C^\adual$. And $\dim D \leq n-k-1$. 
Hence $\wt(D^\adual) \leq (n-k-1)+1 =n-k <n-k+1=\wt(C)$. Hence the subsystem code is
pure. Now assume that $k\leq n-k$. Now it is possible that $C\subseteq C^\adual$. If
$C\subseteq C^\adual$, then $R=1$. So $C\not\subseteq C^\adual$. Now
$\dim D \leq k-1$ from which it follows that $\wt(D^\adual)\leq k \leq n-k<n-k+1=\wt(C)$.
It follows that the subsystem code is pure.
\end{proof}

\section{STABILIZER VERSUS SUBSYSTEM CODES}\label{sec:stabVsoqec}
In this section, we make a rigorous comparison between stabilizer codes
and subsystem codes. Strictly speaking, subsystem codes contain
the class of stabilizer codes; thus, in this section, we assume that the 
subsystem codes have a co-subsystem of dimension greater than 1.

Clearly, there are difficulties in comparing the two classes of codes.
Our ``rain on your parade'' theorem shows that Clifford subsystem
codes cannot have higher distances than stabilizer codes. Their main
edge lies in simpler error recovery schemes. We can quantify this in
terms of the number of syndrome measurements required for
error-correction. This is not necessarily the best method to compare
the decoding complexity. However, it is certainly a reasonable measure if
both codes use table lookup decoding.  In the absence of any special
algorithms for subsystem codes, we will proceed with this as the
metric for comparison.

\subsection{Improving Upon Quantum MDS Codes}
In this subsection, we want to settle whether or not there exist
subsystem code with parameters $[[n,n-2d+2,r>0,d]]_q$.  It turns out
that the bounds that we have derived in Section~\ref{sec:uBounds} will
help in answering this question.  Our best bounds are restricted to
pure codes.  Fortunately, it turns out that all subsystem codes with
parameters $((n,q^{n-2d+2},R,d))_q$ are pure.

\begin{theorem}\label{th:mdsPurity}
Any $((n,q^{n-2d+2},R,d))_q$ Clifford subsystem code is pure.
\end{theorem}
\begin{proof}
If $R=1$, then the claim follows from the fact that quantum MDS codes
are pure, see\cite{rains99}. 

Seeking a contradiction, we assume that there exists an impure
subsystem code with parameters
$((n,q^{n-2d+2},R,d))_q$, exactly pure to $d'<d$ and $R>1$.  
It follows from Corollary~\ref{th:oqec2stab} that it is possible to
construct a stabilizer code with distance $\geq d$ that is (exactly) pure to
$d'$. Then the resulting stabilizer code has the parameters
$((n,q^{n-2d+2},d))_q$ and is impure. But we know that all quantum MDS
codes are pure \cite{rains99}, see also
\cite[Corollary~60]{pre3}. This implies that $d'\geq d$ contradicting
the fact that $d'<d$; hence, every $((n,q^{n-2d+2},R,d))_q$ code is
pure.
\end{proof}

The next theorem explains why Poulin did not have any luck in finding
an $[[5,1,r>0,3]]_2$ subsystem code.
\begin{theorem}
There do not exist any Clifford subsystem codes with parameters
$((n,q^{n-2d+2},R>1,d))_q$. In particular, there do not exist any
$[[n,n-2d+2,r>0,d]]_q$ Clifford subsystem codes.
\end{theorem}
\begin{proof}
Seeking a contradiction, we assume that a subsystem code with
parameters $((n,q^{n-2d+2},R>1,d))_q$ exists. 
By Theorem~\ref{th:mdsPurity}, an $((n,q^{n-2d+2},R,d))_q$ subsystem
code must be pure. It follows from Theorem~\ref{th:pureBound} that a
pure subsystem code with these parameters must satisfy
$$q^{n-2d+2}R\leq q^{n-2d+2}.$$
Therefore, we must have $R=1$, contradicting our assumption $R>1$. 
\end{proof}

\subsection{Better Than Quantum MDS Codes}
In this subsection, we compare once again quantum MDS stabilizer codes
against subsystem codes. We require that both codes are able to encode
the same amount of information and have the same distance. However,
this time, we do not restrict the length of the codes. Our goal is to
determine whether the subsystem code can improve upon an optimal
quantum MDS stabilizer code by fewer syndrome measurements. 

We insist that the codes are $\F_q$-linear, since in this case the
number of syndrome measurements can be directly obtained from the code
parameters.  Indeed, recall that an $\F_q$-linear $[[n,k,r,d]]_q$
subsystem code requires $n-k-r$ syndrome measurements, and an
$\F_q$-linear $[[n',k',d']]_q$ stabilizer code requires $n'-k'$
syndrome measurements. 

\begin{theorem}\label{th:betterQMDS}
If there exists an $\F_q$-linear $[[k+2d-2,k,d]]_q$ quantum MDS
stabilizer code, then an $\F_q$-linear $[[n,k,r,d]]_q$ subsystem code
satisfying 
\begin{equation}\label{eq:length}
k+r\leq n-2d+2
\end{equation} 
cannot require fewer syndrome measurements than the stabilizer code.
\end{theorem}

[We remark that any pure $[[n,k,r,d]]_q$ subsystem code satisfies the
inequality (\ref{eq:length}) by Theorem~\ref{th:pureBound}.]

\begin{proof}
Seeking a contradiction, we assume that the subsystem code requires
fewer syndrome measurements than the quantum MDS code, that is, we
assume that $k+2d-2-k>n-k-r$. This implies that $k+r > n-2d+2$,
contradicting our assumption that $k+r\leq n-2d+2$.
\end{proof}

Now, we can partially answer the question when an 
$\F_q$-linear $[[n,k,r,d]]_q$ subsystem code will lead to
better error recovery schemes than the quantum MDS codes. 

\begin{corollary}
Suppose that an $\F_q$-linear $[[k+2d-2,k,d]]_q$ quantum MDS code $Q$ 
exists. Then an $\F_q$-linear $[[n,k,r,d]]_q$ subsystem code that 
beats the 
the stabilizer code $Q$ must be impure and must satisfy $k+r> n-2d+2$.
\end{corollary}
\begin{proof}
We know from Theorem~\ref{th:pureBound} that all pure $[[n,k,r,d]]_q$ codes satisfy
$k+r\leq n-2d+2$. But Theorem~\ref{th:betterQMDS} implies that such a code
cannot have fewer syndrome measurements than the $\F_q$-linear MDS code. Hence, the
subsystem code, if it is better, must be impure and have $k+r>n-2d+2$.
\end{proof}

\subsection{Better Than Optimal non-MDS Stabilizer Codes}
We know that MDS codes do not exist for all lengths, so it is reasonable to consider
optimal stabilizer codes that are non-MDS. In this case, the comparison is slightly more complicated. An $[[n,k,r,d]]_q$ subsystem code could be better than an
optimal $[[n',k,d]]_q$ stabilizer code. That in itself does not guarantee that the
class of subsystem codes is superior to the class of stabilizer codes. 

For instance, the shortest code to encode $2$ qubits with distance $3$ is $[[8,2,3]]_2$ (see \cite{calderbank98}). Suppose that an $[[8,2,1,3]]_2$ code exists.
This subsystem code requires only $8-2-1=5$ syndrome measurements as
against the $8-2=6$ measurements of the optimal stabilizer code. To
conclude that the subsystem codes are better than stabilizer codes
would be premature, for there exists an $[[8,3,3]]_2$ code
(cf. \cite{calderbank98}) that requires $8-3=5$ syndrome measurements
and encodes one more qubit than the subsystem code. It is therefore
necessary to compare the subsystem code with all optimal $[[n',k'\geq
k ,d]]_q$ stabilizer codes, where $n'$ ranges from $n-r$ to $n$. Only
if the subsystem code requires fewer syndrome measurements in each
case, then we can conclude that the class of subsystem codes leads to
better error recovery schemes.

We do not know precisely the properties of such subsystem codes. For
instance, we do not know if such subsystem code is required to be
impure or if it must satisfy $k+r>n-2d+2$.

Next, we turn our attention to a slightly different 
question which shows that in general whenever good subsystem codes exist, good stabilizer codes also exist.

\subsection{Subsystem Codes and Stabilizer Codes of Comparable
Performance }

The reader will perhaps wonder why one cannot simply discard the gauge
subsystem to obtain a shorter quantum code without sacrificing
distance or dimension.  The reason why we cannot do so is because, in general, 
there is no one to one correspondence between the gauge qudits and the
physical qudits. Yet, our intuition is not entirely misguided as the
following result will show.

\begin{theorem}
Let $\F_q$ be finite field of characteristic $p$ and $1< q^k \leq q^n$ a power of $p$.
Let $r$ be an integer such that $0<r<n$, and 
$$\sum_{j=1}^{d-1} \binom{n}{j}(q^{2}-1)^j (q^{n+k+r}-q^{n+r-k})<(p-1)(q^{2n}-1)$$
holds, then there exist both an
$((n,q^k,q^r,\geq d))_q$ Clifford subsystem code and an 
$((n-r,q^k,\geq d))_q$ stabilizer code. 
\end{theorem}
\begin{proof}
By hypothesis 
$$
\sum_{j=1}^{d-1} \binom{n}{j}(q^{2}-1)^j <(p-1)\frac{(q^{2n}-1)}{(q^{n+k+r}-q^{n+r-k})}$$
holds and Theorem~\ref{th:gvoqec} implies the existence of an
$((n,q^k,q^r,\geq d))_q$ Clifford subsystem code. We can rewrite the RHS 
of the inequality as 
\begin{eqnarray*}
\mbox{RHS}& =&(p-1) \frac{q^{n-r}-q^{-n-r}} {q^k-q^{-k}},\\
&=& (p-1) \frac{q^{n-r}-q^{-n+r}}{ q^k-q^{-k}}+(p-1)\frac{q^{-n+r} -q^{-n-r} }{ q^k-q^{-k}} ,\\
& = & (p-1) \frac{q^{n-r}-q^{-n+r}}{ q^k-q^{-k}}+
\underbrace{(p-1)\frac{q^{-n+r} -q^{-n-r} }{ q^k-q^{-k}}}_{\leq 1,\mbox{ if } r>0}.
\end{eqnarray*}
Now under the assumption $r<n$, we obtain a lower bound for LHS as follows.
\begin{eqnarray*}
\begin{split}
\binom{n}{1}(q^2-1) &+ \sum_{j=2}^{d-1} \binom{n-r}{j}(q^{2}-1)^j 
\\ &=  \sum_{j=1}^{d-1} \binom{n-r}{j}(q^{2}-1)^j +r(q^2-1),\\
 \ & \leq  \sum_{j=1}^{d-1} \binom{n}{j}(q^{2}-1)^j = \mbox{LHS}.
\end{split}
\end{eqnarray*}
Since we know that $\mbox{LHS}<\mbox{RHS}$ we can write 
\begin{eqnarray*}
\begin{split}
\sum_{j=1}^{d-1} &\binom{n-r}{j}(q^{2}-1)^j +\underbrace{r(q^2-1)}_{>1,\mbox{ if } r>0} \\
&< (p-1) \frac{q^{n-r}-q^{-n+r}}{ q^k-q^{-k}}+
\underbrace{(p-1)\frac{q^{-n+r} -q^{-n-r} }{ q^k-q^{-k}}}_{\leq 1, \mbox{ if } r>0},\\
\sum_{j=1}^{d-1} & \binom{n-r}{j}(q^{2}-1)^j  < (p-1) \frac{q^{n-r}-q^{-n+r}}{ q^k-q^{-k}}.
\end{split}
\end{eqnarray*}
Then by Corollary~\ref{th:gvstab}, there exists an $((n-r,q^k,\geq d))_q$ stabilizer code.  
\end{proof}

While they might differ in their distance, the preceding theorem
indicates that in many cases, whenever a good subsystem code exists,
then there will also exist a good stabilizer code encoding as much
information and having comparable distance and of shorter length. The
assumption of integral $r$ may not be much of a restriction in light
of Theorem~\ref{th:gaugeReduction}.

\enlargethispage{-3.3cm}

In comparing the complexity of the error recovery schemes for the two
codes, we run into a small problem since we do not know if the codes
are $\F_q$-linear. Actually, if we use the stronger result of
\cite[Lemma~31]{pre3} and insist that $n\equiv k\bmod 2$, then we can
show that the stabilizer code is $\F_q$-linear. This guarantees that
the stabilizer code will require $n-k-r$ syndrome measurements which
is comparable to that of an $\F_q$-linear subsystem code. It appears
then, quite often, subsystem codes do not offer any gains in error
recovery, as there will exist a corresponding stabilizer code that
encodes as many qudits, of similar distance and equal complexity of
decoding.

\section{CONCLUSION}
In this paper we investigated subsystem codes and their connections to
classical codes.  We derived lower and upper bounds on the parameters
of the subsystem codes. We settled the question whether or not there
exist $[[n,n-2d+2,r>0,d]]_q$ subsystem codes exist.  We showed that
pure $\F_q$-linear subsystem codes do not lead to any reduction in
complexity of error recovery as compared with an $\F_q$-linear MDS
stabilizer code of equal capability. As a consequence we concluded
that the subsystem codes that outperform the quantum MDS codes must be
impure. Further, we showed that quite often the existence of a
subsystem code implies the existence of a stabilizer code of
comparable performance and complexity of error recovery.


\begin{thebibliography}{10}

\bibitem{bacon06}
D.~Bacon.
\newblock Operator quantum error correcting subsystems for self-correcting
  quantum memories.
\newblock {\em Phys. Rev.~A}, 73(012340), 2006.

\bibitem{magma}
W.~Bosma, J.J. Cannon, and C.~Playoust.
\newblock The {M}agma algebra system~{I}: {T}he user language.
\newblock {\em J. Symb. Comp.}, 24:235--266, 1997.

\bibitem{calderbank98}
A.R. Calderbank, E.M. Rains, P.W. Shor, and N.J.A. Sloane.
\newblock Quantum error correction via codes over {GF}(4).
\newblock {\em IEEE Trans. Inform. Theory}, 44:1369--1387, 1998.

\bibitem{grove01}
L.C. Grove.
\newblock {\em Classical Groups and Geometric Algebra}.
\newblock Graduate Studies in Mathematics. American Mathematical Society, 2001.

\bibitem{pre3}
A.~Ketkar, A.~Klappenecker, S.~Kumar, and P.K. Sarvepalli.
\newblock Nonbinary stabilizer codes over finite fields.
\newblock To appear in IEEE Trans. Inform. Theory, November, 2006.

\bibitem{pre7}
A.~Klappenecker and P.~K. Sarvepalli.
\newblock Clifford code constructions of operator quantum error-correcting
  codes.
\newblock 2006.

\bibitem{knill06}
E.~Knill.
\newblock On protected realizations of quantum information.
\newblock Eprint: quant-ph/0603252, 2006.

\bibitem{kribs05c}
D.~W. Kribs.
\newblock A brief introduction to operator quantum error correction.
\newblock Eprint: math/0506491, 2005.

\bibitem{kribs05}
D.~W. Kribs, R.~Laflamme, and D.~Poulin.
\newblock Unified and generalized approach to quantum error correction.
\newblock {\em Phys. Rev. Lett.}, 94(180501), 2005.

\bibitem{kribs05b}
D.~W. Kribs, R.~Laflamme, D.~Poulin, and M.~Lesosky.
\newblock Operator quantum error correction.
\newblock Eprint: quant-ph/0504189, 2005.

\bibitem{poulin05}
D.~Poulin.
\newblock Stabilizer formalism for operator quantum error correction.
\newblock {\em Phys. Rev. Lett.}, 95(230504), 2005.

\bibitem{rains99}
E.M. Rains.
\newblock Nonbinary quantum codes.
\newblock {\em IEEE Trans. Inform. Theory}, 45:1827--1832, 1999.

\end{thebibliography}
\def\cprime{$'$}

\end{document}